\documentstyle[epsfig]{aipproc}

\begin{document}
\title{Circumstellar Nebulae in Young Supernova Remnants}

\author{You-Hua Chu$^*$}
\address{$^*$Astronomy Department, University of Illinois,
1002 W. Green Street, Urbana, IL 61801}

%\lefthead{LEFT head}
%\righthead{RIGHT head}
\maketitle

\begin{abstract}
Supernovae descendent from massive stars explode in media that have 
been modified by their progenitors' mass loss and UV radiation.
The supernova ejecta will first interact with the {\it circumstellar} 
material shed by the progenitors at late evolutionary stages, and then 
interact with the {\it interstellar} material.  Circumstellar nebulae 
in supernova remnants can be diagnosed by their small expansion 
velocities and high [N~II]/H$\alpha$ ratios.  The presence of 
circumstellar nebulae appears ubiquitous among known young supernova 
remnants.  These nebulae can be compared to those around evolved 
massive stars to assess the nature of their supernova progenitors.
Three types of archeological artifacts of supernova progenitors 
have been observed in supernovae and/or young supernova remnants:
(1) deathbed ejecta, (2) circumstellar nebulae, and (3) interstellar 
bubbles.  Examples of these three types are given.
\end{abstract}

\section*{Introduction}

Massive stars end their lives in supernova explosions, but it is not
known exactly at which evolutionary stages massive stars explode.
SN~1987A was the only supernova that had a known progenitor, 
Sk$-$69$^\circ$202 \cite{walborn87}, and its B3I spectral type was 
as shockingly different from theorists' expectation as could be.  
It was eventually recognized that Sk$-$69$^\circ$202 was probably a 
binary, and the companion played a vital role in the stellar mass 
loss and stellar evolution before the supernova explosion \cite{pods92}.  
At present, our knowledge of the transition from massive star to 
supernova is sketchy as best.

Massive stars lose mass throughout their lives in the forms of fast 
and slow winds.  The mass loss rate heightens particularly toward 
the late evolutionary stages.  The ejected stellar material has 
been detected as ``circumstellar nebulae" around Wolf-Rayet (WR) 
stars, luminous blue variables (LBVs), and blue supergiants (BSGs)
\cite{chu91,chu97}.  The circumstellar nebulae around these
stars are visible because these stars have stellar winds to compress
the circumstellar material into dense shells and UV fluxes to ionize
the nebulae.  It is observed that circumstellar nebulae around different
types of stars have different sizes, expansion velocities, and 
abundances, depending on the history of the stellar mass loss \cite{CWG99}.

Circumstellar nebulae can also be detected in young SNRs before they 
are completely shredded beyond recognition by SNR shocks.  The best 
known example of circumstellar material in a young SNR is the 
quasi-stationary flocculi (QSFs) in Cas A \cite{KC77}.  Noting a
similarity in the kinematic properties and [N~II]/H$\alpha$ line 
ratios between these QSFs and the Galactic WR nebula NGC~6888, Kirshner 
and Chevalier \cite{KC77} suggested that the progenitor of Cas A was 
a WN star.  Similarly, if circumstellar nebulae are detected in other
young SNRs, the physical properties of these nebulae can be compared to 
those of known nebulae around massive stars in order to determine the 
nature of the supernova progenitors.  

In this paper, I will first give a brief review on the mass loss from
massive stars and how the mass loss modifies the gaseous environment,
then describe the different types of circumstellar nebulae and
interstellar bubbles that young SNRs interact with, and finally give 
examples of young supernovae or SNRs that show such ``archeological
artifacts" of their massive progenitors.

\section*{Evolution, Mass Loss, and Bubbles of Massive Stars}

The most massive stars, with M$_{\rm ZAMS} >$ 40 M$_\odot$, will
evolve along the sequence O $\rightarrow$ Of 
$\rightarrow$ H-rich WN $\rightarrow$ LBV $\rightarrow$ H-poor WN 
$\rightarrow$ H-free WN $\rightarrow$ WC $\rightarrow$ supernova 
\cite{langer94}.  The less massive stars, on the other hand, evolve 
through the red supergiant (RSG) phase instead of the LBV phase
\cite{HD79}.  Of these different evolutionary stages, fast stellar
winds ($>$ 1,000 km~s$^{-1}$) are seen in main sequence O stars, 
WR stars, and BSGs; moderately fast winds ($<$ 1,000 km~s$^{-1}$)
are seen in LBVs; and slow winds ($\sim$ 20 km~s$^{-1}$) are
seen in RSGs.

These various stellar winds interact with the ambient medium and 
form wind-blown bubbles \cite{weaver77}.  The hydrodynamic 
evolution of these bubbles has been calculated for stars evolving 
along the sequence O star $\rightarrow$ LBV $\rightarrow$ WR star,
e.g., a 60 M$_\odot$ star \cite{GML96}, and for stars evolving along 
the sequence O star $\rightarrow$ RSG $\rightarrow$ WR star, 
e.g., a 35 M$_\odot$ star \cite{GLM96}.

The results of these calculations can be summarized as 
follows.  Massive stars form ``interstellar bubbles" during 
the main sequence stage, as the bubbles consist of mainly
interstellar material.  The copious mass loss during the LBV 
or RSG phase forms a small, slowly-expanding, circumstellar nebula
within the cavity of the interstellar bubble.  During the final WR
phase, the fast stellar wind sweeps up and compresses the 
circumstellar material to form a ``circumstellar bubble," 
which consists of mainly ejected stellar material.  

Interstellar and circumstellar bubbles have both been observed
around WR stars, LBVs, and BSGs \cite{CWG99}.  Interstellar 
and circumstellar bubbles can be easily distinguished according 
to their patterns of expansion and abundances -- circumstellar 
bubbles show regular expansion patterns and high N/O abundance 
ratios.  In general, interstellar bubbles are large, up to a few 
tens of pc in radius.  Circumstellar bubbles of WR stars are 
usually a few pc in radius and expand fast, V$_{\rm exp} \ge$ 
50 km~s$^{-1}$.  Circumstellar bubbles of LBVs are small, 
$\le$ 1 pc in radius; some expand slowly, V$_{\rm exp} <$ 30 
km~s$^{-1}$, while others expand fast with V$_{\rm exp} \sim$
50--100 km~s$^{-1}$.  $\eta$~Car is the most extreme but rare 
case \cite{morse98}.

\section*{Diagnostics of Circumstellar Nebulae}

Circumstellar nebulae consist of material ejected by stars.
Therefore, the nebular expansion is usually regular and shows 
point-symmetry with respect to the central star.  Furthermore,
the elemental abundances show enrichment in CNO products, i.e.,
high N/O ratios \cite{esteban92}.  For normal nebular conditions,
a high N abundance leads to high [N~II]/H$\alpha$ line ratios.
It is thus easy to diagnose a circumstellar nebula using the
expansion pattern and [N~II]/H$\alpha$ ratios.

Circumstellar nebulae in young SNRs can be detected using high
dispersion spectroscopic observations of the H$\alpha$ and
[N~II] $\lambda$6548 lines.  The presence of a narrow emission 
component with anomalously high [N~II]/H$\alpha$ ratio would 
indicate the existence of a circumstellar nebula.

\section*{Circumstellar Nebulae and Interstellar Bubbles in Young SNRs}

Young SNRs contain a rich variety of archeological artifacts left
behind by their massive progenitors.  Three distinct types of 
circumstellar and interstellar nebulae have been observed:
(1) deathbed ejecta, (2) circumstellar nebulae, and (3) interstellar 
bubbles.  Descriptions and examples of these nebulae are given in 
the subsections below.

\subsection*{1. Deathbed Ejecta}

Some supernovae, after the supernova light has faded, show 
narrow emission lines from circumstellar nebulae that are 
characterized by very high densities, 
$\gg$ 10$^6$ H-atom~cm$^{-3}$, and moderate expansion
velocities, $<$~100 km~s$^{-1}$.  The best examples are
SN~1987K \cite{chu99}, SN~1997ab \cite{salamanca98},
and SN~1997eg \cite{salamanca00}.  The H$\alpha$ and [N~II] 
lines of SN~1978K and SN~1997ab are shown in Figure 1.

\begin{figure}[tbh] % fig 1
\vspace{10pt}
\centerline{\epsfig{file=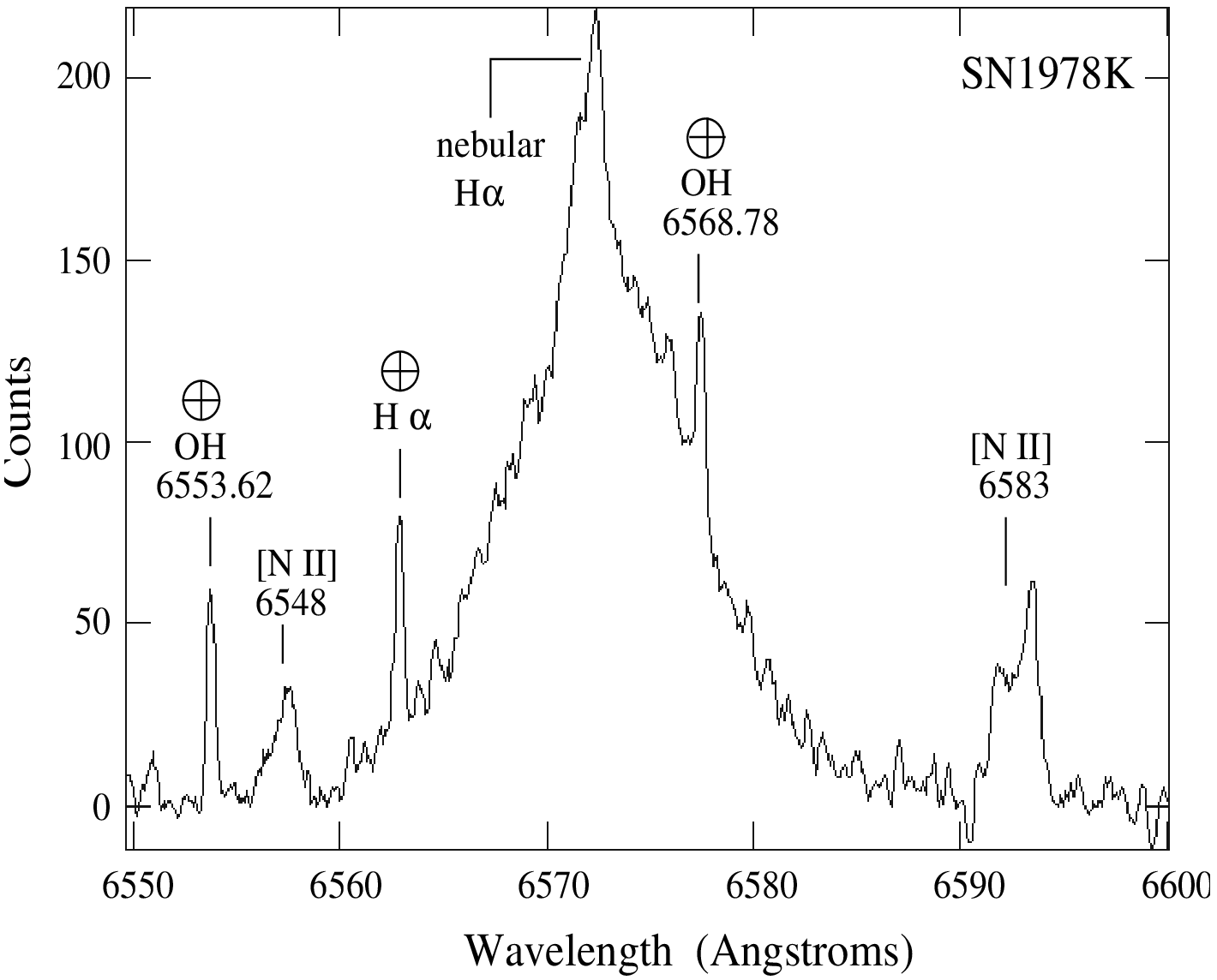,height=1.8in,width=2.7in}
\epsfig{file=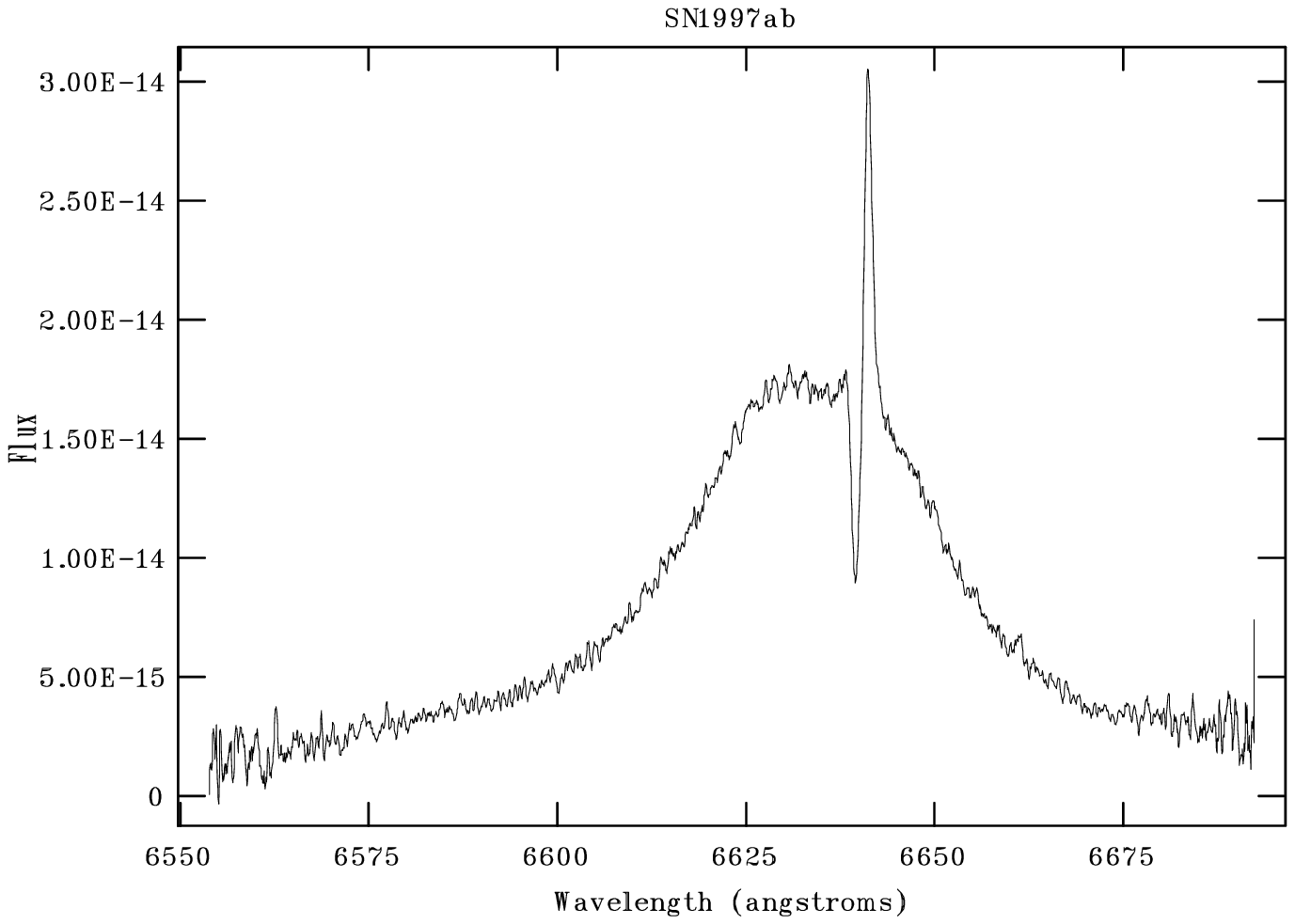,height=1.8in,width=2.7in}}
\vspace{10pt}
\caption{{\bf Left:} High-dispersion H$\alpha$+[N~II] spectrum
SN~1987K, taken from [14]. {\bf Right:} High-dispersion
H$\alpha$ spectrum of SN~1997ab, taken from [15].}
\label{fig1}
\end{figure}

The H$\alpha$+[N~II] spectrum of SN~1978K shows a narrow, nebular 
H$\alpha$ component superposed on the peak of the broad H$\alpha$ 
component of the SN ejecta.  The narrow nebular 
[N~II]$\lambda\lambda$6548,6583 lines are also detected.
The velocity profiles of the nebular components imply an 
expansion velocity $<$ 50 km~s$^{-1}$.  The [N~II]/H$\alpha$ 
ratio is $\sim$ 1.  This nebular emission must originate
from a circumstellar nebula.  Low-dispersion spectra 
suggest a nebular density of a few $\times$10$^5$ 
H-atom cm$^{-3}$ \cite{ryder93}.

The H$\alpha$ spectra of SN~1997ab and SN~1997eg both have a 
narrow, nebular, P Cygni profile superposed on the broad profile 
of the SN ejecta \cite{salamanca98,salamanca00}.  The 
[N~II]/H$\alpha$ ratio of the nebular component is high (Salamanca, 
private communication), and the expansion velocity of the nebular 
component is $<$ 100 km~s$^{-1}$ \cite{salamanca98,salamanca00}, 
suggesting the existence of a circumstellar nebula.  The P Cygni 
profile of the nebular component requires that the density is 
$\ge$ 10$^7$ H-atom cm$^{-3}$.

The densities of the circumstellar nebulae around these three
supernovae are orders of magnitude higher than the densities
observed in circumstellar bubbles of WR stars \cite{chu99}.
The density of SN~1978K's circumstellar nebula is comparable to 
those of the densest LBV nebulae, but the densities of SN~1997ab's
and SN~1997eg's circumstellar nebulae are higher than any known 
circumstellar nebulae.  These circumstellar nebulae, therefore,
must have a different origin from the known circumstellar 
nebulae around evolved massive stars.  They may represent the 
stellar material ejected shortly prior to the supernova explosion,
justifying the name ``deathbed ejecta."

It would be of great interest to determine the elemental abundances
of these circumstellar nebulae, and compare them to the nucleosynthesis
yields of stellar evolution models.  The results will shed light 
on the evolutionary stage of the SN progenitor.  

\subsection*{2. Circumstellar Nebulae}

\subsubsection*{The Case of SN~1987A}

The first example given here is SN~1987A, where we are witnessing
the collision between supernova ejecta and a circumstellar nebula.
HST WFPC2 images of SN~1987A taken after the SN light had faded 
away revealed an inner ring and two outer rings (Figure 2).
The [N~II]/H$\alpha$ ratios of these rings are 4.2 and 2.5 
for the inner and outer rings, respectively \cite{burrow95}.
These high [N~II]/H$\alpha$ ratios suggest anomalous nitrogen abundance, 
indicating that these rings are circumstellar, as opposed to interstellar,
in origin.  The expansion velocity of the inner ring is 
$\sim$ 10 km~s$^{-1}$, indicating that it was not ejected during the SN 
explosion \cite{CK91}.   The morphology and kinematics of these
circumstellar rings place constraints on the SN progenitor's mass loss 
history and lend support to its binarity \cite{pods92}.

\begin{figure}
\centerline{\epsfig{file=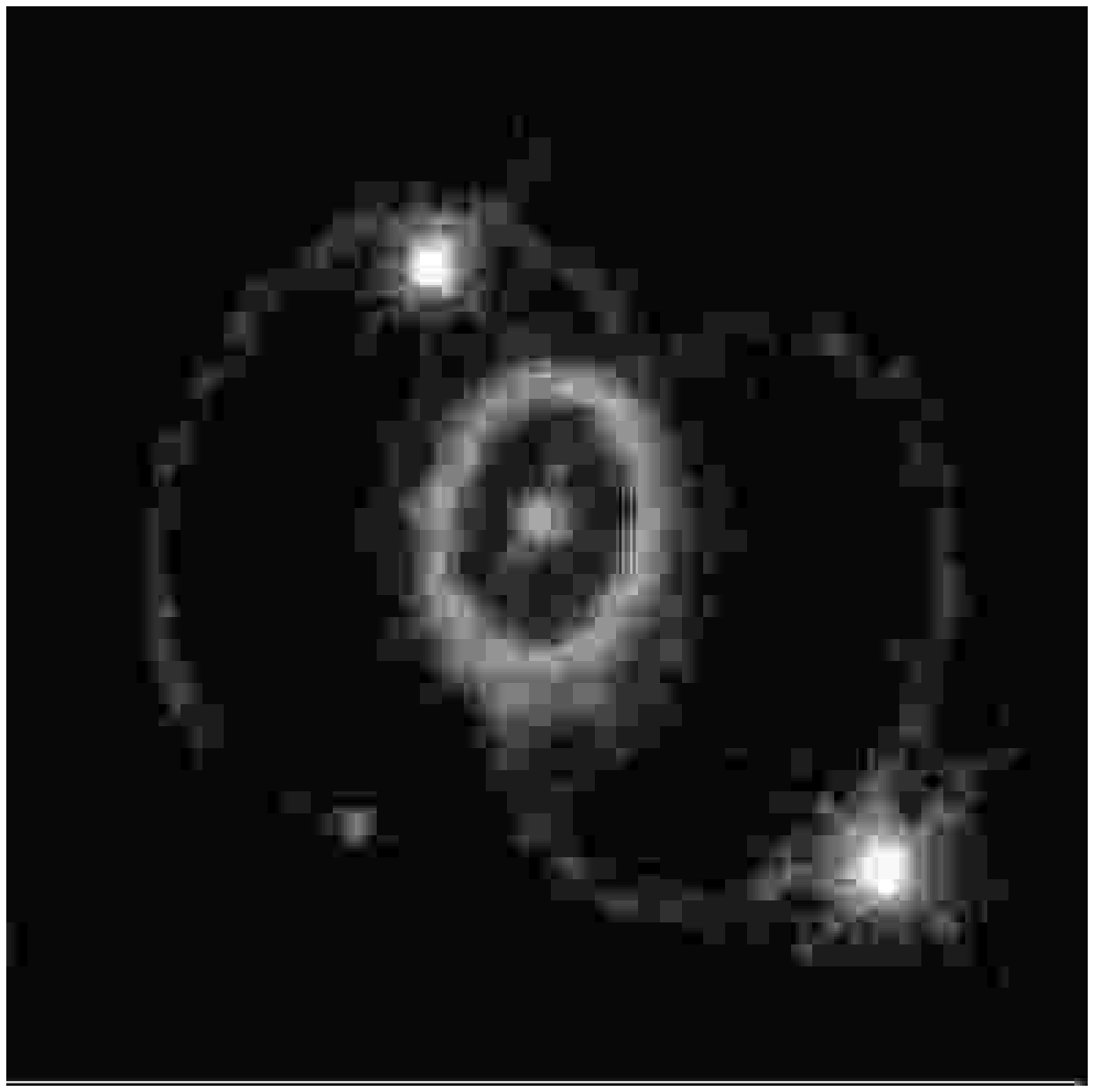,height=2in,width=2in}
\epsfig{file=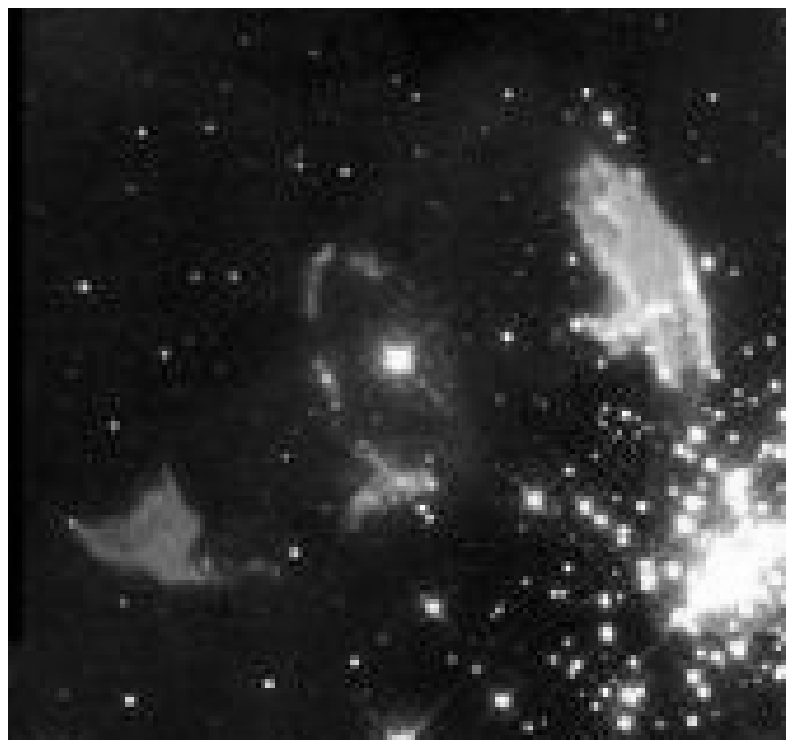,height=1.86in,width=2in}}
\vspace{10pt}
\caption{{\bf Left:} SN~1987A and its circumstellar rings.
   {\bf Right:} Sher 25 and its circumstellar rings. 
    The progenitor of SN~1987A, Sk$-$69~202, was a B3~I star 
    and Sher~25 is a B1.5~I star.  Both inner rings are 
    0.4~pc in diameter!}
\end{figure}

It is interesting to note that circumstellar rings around B
supergiants similar to those of SN~1987A's are probably rare.
Only one case has been serendipitously found around
the B1.5~I star Sher 25 in NGC~3603 (Figure 2, \cite{brandner97b}).
A search for circumstellar nebulae around $>$100 B supergiants in 
the Galaxy and the Magellanic Clouds has yielded null results 
\cite{chu97}.  The similarity in spectral type between Sher 25 
and Sk$-$69~202 and the similarity in the size and physical 
properties of their circumstellar rings suggest that Sher~25 
is a Galactic twin of Sk$-$69~202 and may very well be on the 
verge of a SN explosion \cite{brandner97a}.

\subsubsection*{The Case of SNR~0540$-$69.3}

Next to SN~1987A, the second youngest SNR in the Large Magellanic Cloud
(LMC) is SNR~0540$-$69.3 \cite{caraveo98}.
HST WFPC2 images of SNR 0540$-$69.3 in the H$\alpha$ and
[N~II] lines show a [N~II]-bright ring with a diameter of
$\sim$5$''$, or 1.25 pc (see Figure 3).  Spectroscopic 
observations of the [N~II] lines show narrow velocity profiles 
indicating an expansion velocity $\le$50 km~s$^{-1}$ \cite{caraveo98}.
This [N~II]-bright ring is apparently a circumstellar bubble of the 
SN progenitor.  The size of this circumstellar nebula is comparable to 
those of LBV nebulae, but smaller than WR bubbles \cite{chu99}.  It is
possible that the SN progenitor exploded during or shortly after
an LBV phase.  The elemental abundances of this circumstellar
nebula need to be measured in order to confirm the LBV hypothesis.

\begin{figure}
%\centerline{\epsfig{file=chufig3a.eps,height=2.285in,width=2.8in}
%\epsfig{file=chufig3b.eps,height=2.285in,width=2.8in}}
\vspace{50pt}
\caption{{\bf Left:} HST WFPC2 image of SNR 0540$-$69.3 in
 the [N~II]$\lambda$6583 line.  The circumstellar nebula
 is visible near the center of the PC field.  {\bf Right:} 
 Chandra HRC X-ray image of SNR 0540$-$69.3.  The X-ray 
 emission peaks at the region of circumstellar nebula.
 These two images have the same image scales.}
\vspace{30pt}
\centerline{\epsfig{file=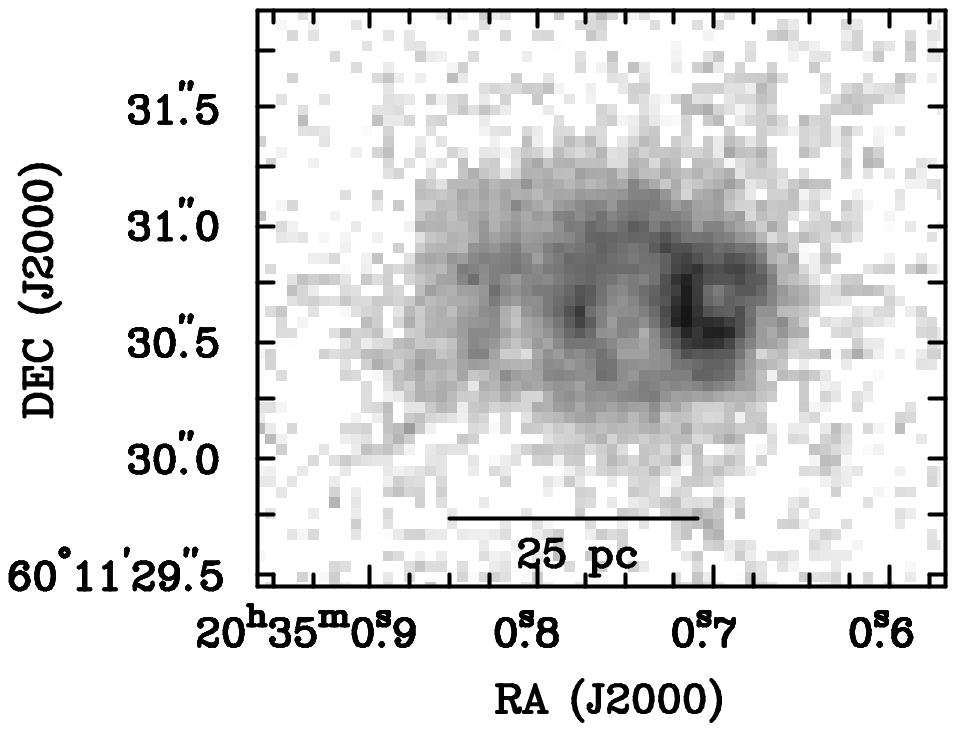,height=2.33in,width=2.8in}
\epsfig{file=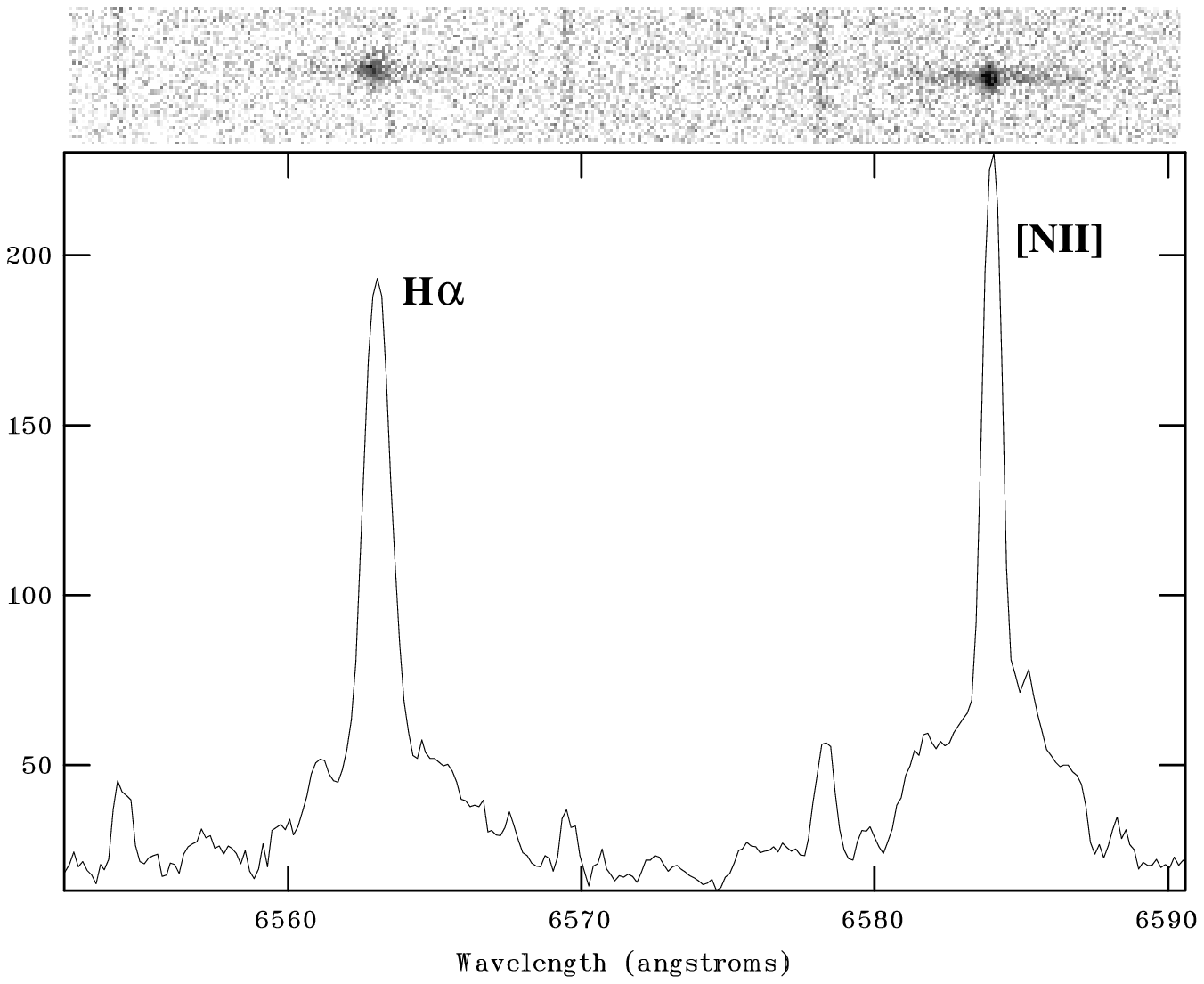,height=2.33in,width=2.8in}}
\vspace{10pt}
\caption{{\bf Left:} HST WFPC2 H$\alpha$ image of MF~16 in NGC~6946.
  {\bf Right:} Echellogram and profile plot of the H$\alpha$+[N~II] 
  lines of MF~16 in NGC~6946.  Both figures are taken from [28].}
\end{figure}

\subsubsection*{The Case of SNR MF16 in NGC~6946}

The SNR MF~16 in NGC~6946 \cite{MF97} at a distance of $\sim$5 Mpc is 
known for its very high X-ray and optical luminosities, about 1,000 
times as luminous as Cas A in X-rays \cite{schlegel94,BF94}.
HST WFPC2 images of MF~16 (Figure 4) show a multiple-loop morphology, 
which leads to the suggestion that the high luminosity of this remnant 
is caused by colliding SNRs. \cite{blair00}.  However. a
different interpretation based on ground-based high-dispersion 
spectroscopic observations of MF~16 has been suggested \cite{dunne00}.
The spectrum of MF~16 shows narrow H$\alpha$ and [N~II] lines superposed 
on the broad SNR components (Figure 4), and the [N~II]/H$\alpha$ ratio 
for the narrow component is high, $\sim$1, indicating that the narrow 
components are emitted by N-enriched material.  The expansion velocity 
implied by the velocity widths is $<$15 km~s$^{-1}$.  It is thus possible 
that the narrow components originate in a circumstellar bubble and
the collision between this circumstellar bubble and the SNR causes
the high X-ray and optical luminosities.  HST STIS long-slit observations
of MF~16 are needed to examine the location of regions with high 
[N~II]/H$\alpha$ ratios in order to determine the boundary of the
circumstellar bubble.  Ground-based, kinematically-resolved 
spectrophotometric observations similar to those of NGC~6888
\cite{EV92} are needed to determine the abundances of this 
circumstellar material. 

\subsection*{3. Interstellar Bubbles}

Supernova ejecta in young SNRs, once having swept past the circumstellar
material, can advance quickly to the inner walls of the interstellar bubble
formed by the progenitor at the main sequence stage.  The SNR N132D in the
LMC is a good example \cite{morse96}.  The combination of high X-ray 
brightnesses and large linear sizes is usually indicative of such young SNRs.
The detection/confirmation of an interstellar bubble is non-trivial, as
illustrated in the example of N63A.

N63A is a SNR associated with the OB association LH83 in the 
LMC (Figure 5). Its optical image shows a three-lobed, 
clover-shaped nebulosity about 20$''$ ($\sim$ 5 pc) across.
The two eastern lobes exhibit a high [S II]/H$\alpha$ ratio, 
indicating a shock excitation, while the western lobe shows 
spectra characteristic of photoionization.  The fact that
N63A has such a high X-ray surface brightness and the 
fact that the optical size of the SNR N63A is much smaller 
than the extent of radio and X-ray emission ($\sim$70$''$)
suggest that the SNR N63A is young and its SNR shock has just 
reached the inner walls of the progenitor's interstellar bubble.

\begin{figure}
%\centerline{\epsfig{file=chufig5a.eps,height=2.8in,width=2.8in}
%\epsfig{file=chufig5b.eps,height=2.8in,width=2.8in}}
\vspace{10pt}
%\centerline{\epsfig{file=chufig5c.eps,height=2in,width=2in}
\hspace{30pt}
\vspace{10pt}
\caption{{\bf Upper Left:} H$\alpha$ image of N63, overlaid by
X-ray contours extracted from a ROSAT HRI observation.  This
image shows that the SNR is associated with an OB association
and embedded in an HII region. {\bf Upper Right:} HST WFPC2
image of N63A in H$\alpha$.  This image resolves not only the
three bright lobes but also small cloudlets within the X-ray
emission regions.  {\bf Lower Left:} Close-up of a small evaporating
cloudlet.  {\bf Lower Right:} HST WFPC2 image of N63A in [O III].
The southern boundary of the SNR is delineated by a faint [O III]
arc which is coincident with the X-ray boundary, suggesting that the
[O~III] arc marks the interstellar bubble blown by the supernova
progenitor.}
\end{figure}

HST WFPC2 images of N63A in H$\alpha$ and [S~II] resolved the 
filamentary structure in the two eastern lobes and diffuse 
emission in the western lobe, consistent with the excitation 
mechanisms diagnosed from their spectral properties \cite{chuiau99}.  
Additionally, the WFPC2 images reveal a number of cloudlets as 
small as 0.1 pc across, within the X-ray-emitting regions of the SNR.
The [S II]/H$\alpha$ ratios and locations of these cloudlets suggest
that these are shocked cloudlets lagging behind the shock front.
These evaporating cloudlets are probably responsible for injecting
mass into the hot SNR interior to produce the high X-ray surface 
brightness.  The H$\alpha$ and [S~II] images fail to show the
location of the interstellar bubble.

A recent HST WFPC2 image of N63A in the [O~III] $\lambda$5007 line
finally shows a faint arc at the southern boundary of the SNR.
The location of the arc coincides with the X-ray boundary.  
The long sought interstellar bubble of N63A is finally found.

\section*{Summary and Conclusion}

Young SNRs offer the best laboratories to study the archeological 
artifacts of massive supernova progenitors.  It is possible to
detect the very latest stellar ejecta (the deathbed ejecta), 
circumstellar nebulae ejected at late evolutionary stages, and
interstellar bubbles blown by the progenitors at main sequence
stage.  Detailed spectroscopic observations of these nebulae
are necessary to determine their abundances in order to assess
the evolutionary stages of the progenitors right before the 
supernova explosion.  

\vspace{20pt}
This research is partially supported the the HST grant
STScIGO-08110.01-97A.

\end{document}